\documentclass[11pt]{article}

\usepackage{adjustbox}
\usepackage{float}
\usepackage[T1]{fontenc}
\usepackage{graphicx}
\usepackage{hhline}
\usepackage[utf8]{inputenc}
\usepackage{multicol}
\usepackage{multirow}
\usepackage[paperheight=29.7cm,paperwidth=21.0cm,left=3.0cm,right=3.0cm,top=2.5cm,bottom=2.5cm]{geometry}
\usepackage{xurl}

\begin{document}

\vspace{1\baselineskip}
\begin{center}
\textbf{The Río Hortega University Hospital Glioblastoma dataset: a comprehensive collection of preoperative, early postoperative and recurrence MRI scans (RHUH-GBM)}
\end{center}

\vspace{1\baselineskip}
\begin{center}
\textit{Santiago Cepeda\textsuperscript{1}, Sergio García-García\textsuperscript{1}, Ignacio Arrese\textsuperscript{1}, Francisco Herrero\textsuperscript{2}, Trinidad Escudero\textsuperscript{2}, Tomás Zamora\textsuperscript{3}, Rosario Sarabia\textsuperscript{1}}
\end{center}

\vspace{1\baselineskip}
{\small \textsuperscript{1} Department of Neurosurgery, Río Hortega University Hospital, Dulzaina 2, 47012 Valladolid, Spain}

{\small \textsuperscript{2} Department of Radiology, Río Hortega University Hospital, Dulzaina 2, 47012 Valladolid, Spain}

{\small \textsuperscript{3} Department of Pathology, Río Hortega University Hospital, Dulzaina 2, 47012 Valladolid, Spain}

\vspace{1\baselineskip}
\textbf{ABSTRACT:}

Glioblastoma, a highly aggressive primary brain tumor, is associated with poor patient outcomes. Although magnetic resonance imaging (MRI) plays a critical role in diagnosing, characterizing, and forecasting glioblastoma progression, public MRI repositories present significant drawbacks, including insufficient postoperative and follow-up studies as well as expert tumor segmentations. To address these issues, we present the "Río Hortega University Hospital Glioblastoma Dataset (RHUH-GBM)," a collection of multiparametric MRI images, volumetric assessments, molecular data, and survival details for glioblastoma patients who underwent total or near-total enhancing tumor resection. The dataset features expert-corrected segmentations of tumor subregions, offering valuable ground truth data for developing algorithms for postoperative and follow-up MRI scans.

The public release of the RHUH-GBM dataset significantly contributes to glioblastoma research, enabling the scientific community to study recurrence patterns and develop new diagnostic and prognostic models. This may result in more personalized, effective treatments and ultimately improved patient outcomes.

\vspace{1\baselineskip}

\vspace{1\baselineskip}
\textbf{INTRODUCTION}

Glioblastoma, the most prevalent primary brain tumor, carries a dismal prognosis despite the extensive efforts of clinical trials and research investigations. In recent years, the integration of advanced medical image processing techniques with artificial intelligence-based algorithms has catalyzed the quest to optimize diagnostic accuracy and prognostic models. A key data source leveraged in this domain is magnetic resonance imaging (MRI). Public MRI repositories, such as The Cancer Imaging Archive (TCIA) \textsuperscript{1} and the Multimodal Brain Tumor Segmentation (BraTS) challenge dataset \textsuperscript{2–4}, have supplied a wealth of data for research in this field.

Nonetheless, these public resources have predominantly focused on preoperative studies,  encompassing patients with diverse tumor extent of resection (EOR) categories and limited expert annotations, while primarily featuring structural sequences of MRI modalities. To address these constraints, we introduce the ``Río Hortega University Hospital Glioblastoma Dataset (RHUH-GBM)", comprising multiparametric structural and diffusion MRI images captured at three critical junctures: preoperative, early postoperative (within 72 hours), and follow-up examinations upon recurrence diagnosis. Furthermore, the dataset exclusively contains patients who have undergone total or near-total resection of the enhancing tumor, along with tumor subregion segmentations for each time point. Complementing this data are clinical information, volumetric assessments of resection extents, molecular data, and survival details.

By publicly disseminating this dataset, we aim to enable the scientific community to scrutinize recurrence patterns in patients who have experienced gross total or near-total resection, and facilitate the development of novel registration and segmentation algorithms tailored for postoperative and follow-up MRI scans.

\vspace{1\baselineskip}
\textbf{METHODS}

\textbf{Patient population}

The dataset comprises consecutive patients who underwent surgery between January 2018 and December 2022, with a confirmed histopathological diagnosis of WHO grade 4 astrocytoma. Forty patients were selected based on the following inclusion criteria: 1) Gross total resections (GTR) or Near Total Resection (NTR), defined as having no residual tumor enhancement and an extent of resection exceeding 95$\%$ of the initial enhancing volume, respectively \textsuperscript{5,6}. 2) Availability of MRI studies at three time points: preoperative, early postoperative (within 72 hours), and follow-up studies during which recurrence was diagnosed. 3) Availability of structural T1-weighted (T1w), T2-weighted (T2w), T1 contrast-enhanced (T1ce), Fluid-attenuated inversion recovery (FLAIR), and diffusion-weighted imaging-derived apparent diffusion coefficient (ADC) maps for each study. 4) Receipt of adjuvant treatment with chemotherapy and radiotherapy following the Stupp protocol \textsuperscript{7}.

Patients with severe image acquisition artifacts or missing MRI series were excluded. The modified Response Assessment in Neuro-Oncology (RANO) criteria were utilized to determine tumor progression \textsuperscript{8}. This study received approval from the Institutional Review Board of the Río Hortega University Hospital and the Ethics Committee for Drug Research (CEIm) of the West Valladolid Health Area (Ref. 22PI-208)

\vspace{1\baselineskip}
\textbf{Clinical, Pathological, and Imaging Data}

Clinical and pathological information was obtained from electronic medical records, including age, sex, histopathological diagnosis, pre- and postoperative Karnofsky Performance Score (KPS), isocitrate dehydrogenase (IDH) status, use of operative adjuncts, volumetric assessment of the extent of resection of the contrast-enhancing and non-enhancing tumor, presence of postoperative neurological deficits, details of chemotherapy and radiotherapy received, and overall survival and progression-free survival times. In this collection, all postoperative MRI scans were conducted at Río Hortega University Hospital. Out of the total sample, a subset of 11 patients had initially undergone preoperative and subsequent follow-up MRI scans at a secondary healthcare facility before being referred to the primary center. Details of the MR imaging acquisition parameters are described in Supplementary Table 1.

\vspace{1\baselineskip}
\textbf{Image Preprocessing}

Images were retrieved from the Picture Archiving Communication System (PACS) in Digital Imaging and Communications in Medicine (DICOM) format for subsequent processing. The first step involved converting the images to Neuroimaging Informatics Technology Initiative (NIfTI) format using the dicom2niix tool version v1.0.20220720 \textsuperscript{9}, available at \url{https://github.com/rordenlab/dcm2niix/releases/tag/v1.0.20220720}. Subsequently, the T1ce scans for each subject were registered to the SRI24 anatomical atlas space \textsuperscript{10} using the FLIRT (FMRIB's Linear Image Registration Tool) \textsuperscript{11,12} available at \url{https://fsl.fmrib.ox.ac.uk/fsl/fslwiki/FSL}. The T1w, T2w, FLAIR scans, and ADC maps were then registered to the transformed T1ce scan, resulting in co-registered resampled volumes of 1 $\times$ 1 $\times$ 1 mm isotropic voxels. The brain was extracted from all co-registered scans using a deep learning tool called Synthstrip \textsuperscript{13}, included in FreeSurfer v7.3.0, available at \url{https://github.com/freesurfer/freesurfer/tree/dev/mri_synthstrip}.

Finally, intensity Z-scoring normalization was performed using the normalization tools included in Cancer Imaging Phenomics Toolkit (CaPTk) v1.9.0 \textsuperscript{14} available at \url{https://www.nitrc.org/projects/captk/}

\vspace{1\baselineskip}
\textbf{Tumor Subregions Segmentations}

The preprocessed images from each time point were used as input for generating computer-aided segmentations using Deep-Medic \textsuperscript{15}. Three labels were subsequently obtained, corresponding to 1 - necrosis, 2 - peritumoral signal alteration, including edema and non-enhancing tumor, and 3 - enhancing tumor. All segmentations were carefully reviewed and manually corrected by two expert neurosurgeons specializing in neuroimaging (S.C. and S.G.), who have a solid background in such tasks. 

\vspace{1\baselineskip}
\begin{figure}[H]
\includegraphics[width=14.98cm,height=15.0cm]{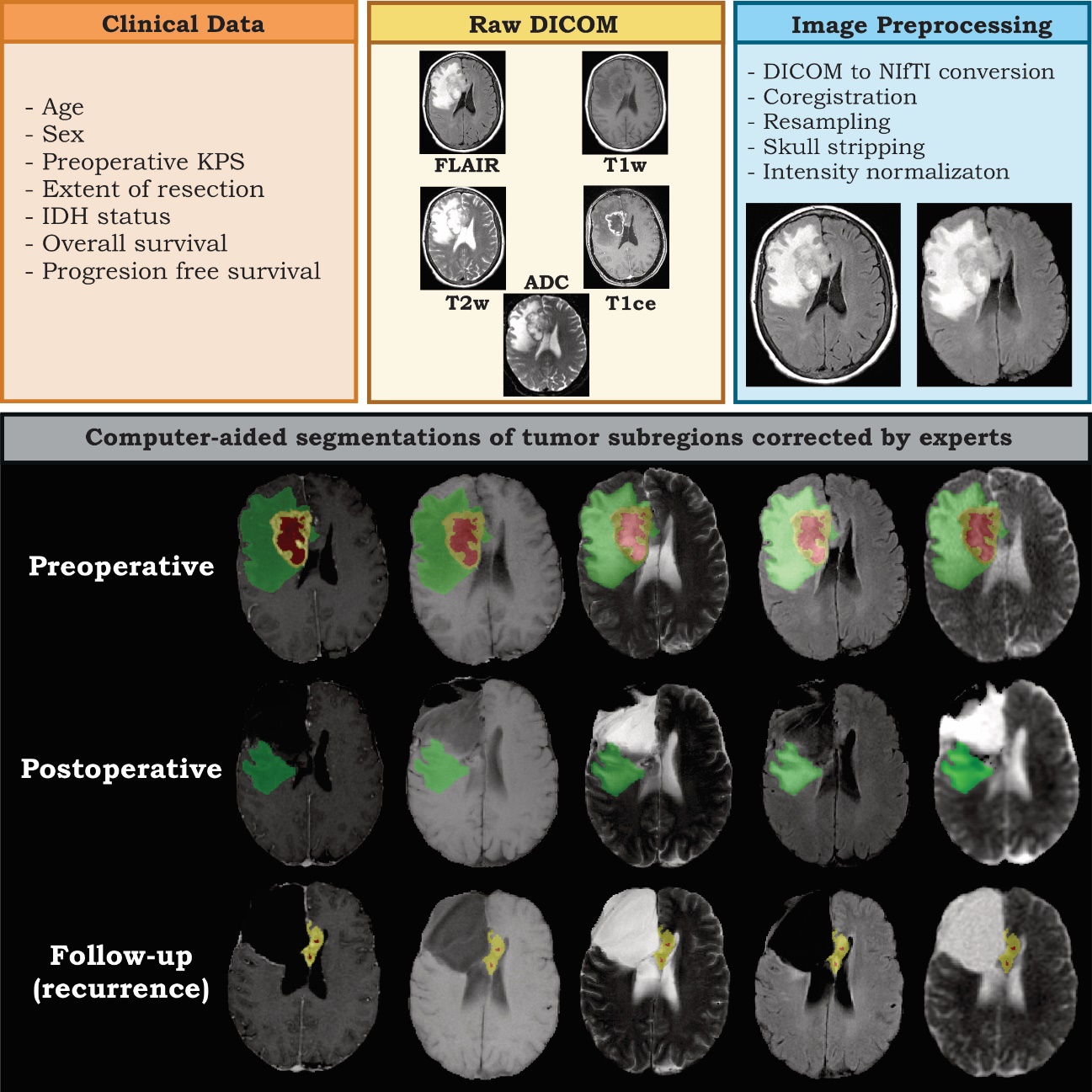}
\end{figure}

\vspace{1\baselineskip}
\textbf{RESULTS}

A summary of the demographic data is presented in Table 1. The patients had an average age of 63 $\pm$ 9 years, consisting of 28 men (70$\%$) and 12 women (30$\%$). The median preoperative Karnofsky Performance Scale (KPS) score was 80. Out of the 40 patients, 38 (95$\%$) were diagnosed with de novo glioblastomas, while two patients (5$\%$) had recurrent glioblastomas previously treated with standard chemoradiotherapy. Four cases (10$\%$) were IDH-mutated, and 36 cases (90$\%$) were IDH wild-type.

The mean preoperative contrast-enhancing tumor volume was 34.99 $\pm$ 26.59, and the mean postoperative contrast-enhancing residual tumor volume was 0.23 $\pm$ 0.47. The mean preoperative T2/FLAIR abnormality was 73.14 $\pm$ 43.63, while the mean postoperative T2/FLAIR abnormality was 35.00 $\pm$ 26.74. Among the patients, 27 (67.5$\%$) underwent gross total resection, and 13 (32.5$\%$) underwent near-total resection. The mean extent of resection (EOR) was 99.31 $\pm$ 1.36$\%$. The median overall survival was 364 days, and the median progression-free survival was 198 days.

\begin{table}[H]
\begin{adjustbox}{max width=\textwidth}
\begin{tabular}{p{7.49cm}p{5.18cm}p{2.31cm}}
\hline
\multicolumn{3}{|p{14.98cm}|}{{\footnotesize \textbf{Table 1. Study population demographics of the Río Hortega University Hospital Glioblastoma dataset (RHUH-GBM)}\par}} \\ 
\hline
\multicolumn{1}{|p{7.49cm}}{{\footnotesize \textbf{Sex}}} & 
\multicolumn{1}{|p{5.18cm}}{\centering
{\footnotesize Male}} & 
\multicolumn{1}{|p{2.31cm}|}{\centering
{\footnotesize 28 (70$\%$)}} \\ 
\hline
\multicolumn{1}{|p{7.49cm}}{} & 
\multicolumn{1}{|p{5.18cm}}{\centering
{\footnotesize Female}} & 
\multicolumn{1}{|p{2.31cm}|}{\centering
{\footnotesize 12 (30$\%$)}} \\ 
\hline
\multicolumn{1}{|p{7.49cm}}{{\footnotesize \textbf{Age (years)}}} & 
\multicolumn{2}{|p{7.49cm}|}{\centering
{\footnotesize 63 $\pm$ 9}} \\ 
\hline
\multicolumn{1}{|p{7.49cm}}{{\footnotesize \textbf{Extent of resection}}} & 
\multicolumn{1}{|p{5.18cm}}{\centering
{\footnotesize GTR}} & 
\multicolumn{1}{|p{2.31cm}|}{\centering
{\footnotesize 27 (67.5$\%$)}} \\ 
\hline
\multicolumn{1}{|p{7.49cm}}{} & 
\multicolumn{1}{|p{5.18cm}}{\centering
{\footnotesize NTR}} & 
\multicolumn{1}{|p{2.31cm}|}{\centering
{\footnotesize 13 (32.5$\%$)}} \\ 
\hline
\multicolumn{1}{|p{7.49cm}}{{\footnotesize \textbf{Number of time-point MRI studies}}} & 
\multicolumn{2}{|p{7.49cm}|}{\centering
{\footnotesize 120}} \\ 
\hline
\multicolumn{1}{|p{7.49cm}}{{\footnotesize \textbf{Number of MRI series}}} & 
\multicolumn{2}{|p{7.49cm}|}{\centering
{\footnotesize 600}} \\ 
\hline
\multicolumn{1}{|p{7.49cm}}{{\footnotesize \textbf{IDH status}}} & 
\multicolumn{1}{|p{5.18cm}}{\centering
{\footnotesize Mutant}} & 
\multicolumn{1}{|p{2.31cm}|}{\centering
{\footnotesize 4 (10$\%$)}} \\ 
\hline
\multicolumn{1}{|p{7.49cm}}{} & 
\multicolumn{1}{|p{5.18cm}}{\centering
{\footnotesize Wild type}} & 
\multicolumn{1}{|p{2.31cm}|}{\centering
{\footnotesize 36 (90$\%$)}} \\ 
\hline
\multicolumn{1}{|p{7.49cm}}{{\footnotesize \textbf{Preoperative KPS}}} & 
\multicolumn{2}{|p{7.49cm}|}{\centering
{\footnotesize 80 (10)}} \\ 
\hline
\multicolumn{1}{|p{7.49cm}}{{\footnotesize \textbf{Operative adjuncts}}} & 
\multicolumn{1}{|p{5.18cm}}{\centering
{\footnotesize 5’ALA}} & 
\multicolumn{1}{|p{2.31cm}|}{\centering
{\footnotesize 40 (100$\%$)}} \\ 
\hline
\multicolumn{1}{|p{7.49cm}}{} & 
\multicolumn{1}{|p{5.18cm}}{\centering
{\footnotesize Sodium Fluorescein}} & 
\multicolumn{1}{|p{2.31cm}|}{\centering
{\footnotesize 7 (17.5$\%$)}} \\ 
\hline
\multicolumn{1}{|p{7.49cm}}{} & 
\multicolumn{1}{|p{5.18cm}}{\centering
{\footnotesize Neuronavigation}} & 
\multicolumn{1}{|p{2.31cm}|}{\centering
{\footnotesize 40 (100$\%$)}} \\ 
\hline
\multicolumn{1}{|p{7.49cm}}{} & 
\multicolumn{1}{|p{5.18cm}}{\centering
{\footnotesize IoUS}} & 
\multicolumn{1}{|p{2.31cm}|}{\centering
{\footnotesize 40 (100$\%$)}} \\ 
\hline
\multicolumn{1}{|p{7.49cm}}{} & 
\multicolumn{1}{|p{5.18cm}}{\centering
{\footnotesize IONM}} & 
\multicolumn{1}{|p{2.31cm}|}{\centering
{\footnotesize 4 (10$\%$)}} \\ 
\hline
\multicolumn{1}{|p{7.49cm}}{} & 
\multicolumn{1}{|p{5.18cm}}{\centering
{\footnotesize DES}} & 
\multicolumn{1}{|p{2.31cm}|}{\centering
{\footnotesize 3 (7.5$\%$)}} \\ 
\hline
\multicolumn{1}{|p{7.49cm}}{{\footnotesize \textbf{Preoperative contrast enhancing tumor volume}}} & 
\multicolumn{2}{|p{7.49cm}|}{\centering
{\footnotesize 34.99 $\pm$ 26.59}} \\ 
\hline
\multicolumn{1}{|p{7.49cm}}{{\footnotesize \textbf{Preoperative T2/FLAIR peritumoral signal alteration volume}}} & 
\multicolumn{2}{|p{7.49cm}|}{\centering
{\footnotesize 35.00 $\pm$ 26.74}} \\ 
\hline
\multicolumn{1}{|p{7.49cm}}{{\footnotesize \textbf{Postoperative contrast enhancing tumor volume}}} & 
\multicolumn{2}{|p{7.49cm}|}{\centering
{\footnotesize 0.23 $\pm$ 0.47}} \\ 
\hline
\multicolumn{1}{|p{7.49cm}}{{\footnotesize \textbf{Postoperative T2/FLAIR peritumoral signal alteration volume}}} & 
\multicolumn{2}{|p{7.49cm}|}{\centering
{\footnotesize 35.00 $\pm$ 26.74}} \\ 
\hline
\multicolumn{1}{|p{7.49cm}}{{\footnotesize \textbf{Radiotherapy treatment details}}} & 
\multicolumn{1}{|p{5.18cm}}{\centering
{\footnotesize VMAT-IMRT-IGRT/60 Gy /30 fx}} & 
\multicolumn{1}{|p{2.31cm}|}{\centering
{\footnotesize 29 (72.5$\%$)}} \\ 
\hline
\multicolumn{1}{|p{7.49cm}}{} & 
\multicolumn{1}{|p{5.18cm}}{\centering
{\footnotesize VMAT-IMRT-IGRT/50 Gy /20 fx}} & 
\multicolumn{1}{|p{2.31cm}|}{\centering
{\footnotesize 6 (15$\%$)}} \\ 
\hline
\multicolumn{1}{|p{7.49cm}}{} & 
\multicolumn{1}{|p{5.18cm}}{\centering
{\footnotesize VMAT-IMRT-IGRT/40.5 Gy /15 fx}} & 
\multicolumn{1}{|p{2.31cm}|}{\centering
{\footnotesize 5 (12.5$\%$)}} \\ 
\hline
\multicolumn{1}{|p{7.49cm}}{{\footnotesize \textbf{Postoperative neurological deficit}}} & 
\multicolumn{1}{|p{5.18cm}}{\centering
{\footnotesize No}} & 
\multicolumn{1}{|p{2.31cm}|}{\centering
{\footnotesize 26 (65$\%$)}} \\ 
\hline
\multicolumn{1}{|p{7.49cm}}{} & 
\multicolumn{1}{|p{5.18cm}}{\centering
{\footnotesize Transient}} & 
\multicolumn{1}{|p{2.31cm}|}{\centering
{\footnotesize 6 (15$\%$)}} \\ 
\hline
\multicolumn{1}{|p{7.49cm}}{} & 
\multicolumn{1}{|p{5.18cm}}{\centering
{\footnotesize Minor persistent}} & 
\multicolumn{1}{|p{2.31cm}|}{\centering
{\footnotesize 6 (15$\%$)}} \\ 
\hline
\multicolumn{1}{|p{7.49cm}}{} & 
\multicolumn{1}{|p{5.18cm}}{\centering
{\footnotesize Major persistent}} & 
\multicolumn{1}{|p{2.31cm}|}{\centering
{\footnotesize 2 (5$\%$)}} \\ 
\hline
\multicolumn{1}{|p{7.49cm}}{{\footnotesize \textbf{Posotperative KPS}}} & 
\multicolumn{2}{|p{7.49cm}|}{\centering
{\footnotesize 80 (20)}} \\ 
\hline
\multicolumn{3}{|p{14.98cm}|}{{\footnotesize Numerical values are expressed in mean and standard deviation or median and interquartile range accordingly. GTR $=$ gross total resection, NTR $=$ near total resection, IDH $=$ isocitrate dehydrogenase, VMAT $=$ Volumetric Modulated Arc Therapy, IMRT $=$ Intensity Modulated Radiation Therapy, IGRT $=$ Image-Guided Radiation Therapy. Radiotherapy treatments are expressed in dose (Gy) and number of fractions (fx)\par}} \\ 
\hline
\end{tabular}
\end{adjustbox}
\end{table}
\vspace{14\baselineskip}
\textbf{DATA AVAILABILITY}

The RHUH-GBM dataset will be publicly available via The Cancer Imaging Archive web site \url{https://www.cancerimagingarchive.net/}. The code applied for image preprocessing is available to the public through a GitHub repository \url{https://github.com/smcch/RHUH-GBM-dataset-MRI-preprocessing}

\vspace{1\baselineskip}
\textbf{DISCUSSION}

Numerous public MRI collections, comprising glioblastoma patients, are available for researchers investigating this aggressive form of brain cancer. These collections have significantly contributed to advancements in the field. Notable datasets include The Cancer Genome Atlas Glioblastoma Multiforme (TCGA-GBM) \textsuperscript{16} , Clinical Proteomic Tumor Analysis Consortium Glioblastoma (CPTAC-GBM) \textsuperscript{17}, Quantitative Imaging Network Glioblastoma (QIN GBM) \textsuperscript{18}, American College of Radiology Imaging Network Fluoromisonidazole-Brain (ACRIN-FMISO-Brain) \textsuperscript{19,20}, and Brain Tumor Segmentation (BraTS) Challenge 2021 \textsuperscript{2–4} . While these collections are invaluable, they primarily focus on preoperative studies and often lack information on the extent of tumor resection performed. Furthermore, only the BraTS dataset includes tumor segmentations.

The Ivy Glioblastoma Atlas Project (Ivy GAP) \textsuperscript{21} collection offers longitudinal studies for 39 patients, but not all include early postoperative assessments. Additionally, this collection features patients with subtotal resections and does not provide segmentations.

The recent Multi-parametric magnetic resonance imaging (mpMRI) scans for de novo Glioblastoma (GBM) patients from the University of Pennsylvania Health System (UPENN-GBM) \textsuperscript{22} and The University of California San Francisco Preoperative Diffuse Glioma MRI (UCSF-PDGM) \textsuperscript{23} collections contain 611 and 501 cases, respectively. These datasets include expert-refined segmentations and additional perfusion and diffusion modalities. However, the UPENN-GBM dataset has a limited 8.9$\%$ of patients with follow-up studies and lacks early postoperative MRI scans. Meanwhile, the UCSF-PDGM dataset provides only preoperative studies for the 402 grade 4 astrocytoma cases.

The HURH-GBM collection enhances the available resources by incorporating early postoperative studies and recurrence scans. The dataset's expert-corrected segmentations are especially advantageous for postoperative scans, given the labor-intensive and time-consuming nature of differentiating peritumoral T2-FLAIR alterations from postoperative changes such as edema, hemorrhage, and ischemia. The inclusion of patients with gross total resection in this collection may potentially facilitate investigations into tumor recurrence patterns. Additionally, the provided labels could act as ground truths for the development of segmentation and coregistration algorithms targeting postoperative and follow-up studies.

\vspace{2\baselineskip}
\textbf{ACKNOWLEDGEMENTS}

This work was partially funded by a grant awarded by the ``Instituto Carlos III, Proyectos I-D-i , Acción Estratégica en Salud 2022", under the project titled "Prediction of tumor recurrence in glioblastomas using magnetic resonance imaging, machine learning, and transcriptomic analysis: A supratotal resection guided by artificial intelligence," reference PI22/01680.

\vspace{1\baselineskip}
\textbf{REFERENCES}

1.\ \ \ \ Clark K, Vendt B, Smith K, et al. The Cancer Imaging Archive (TCIA): maintaining and operating a public information repository. \textit{J Digit Imaging}. 2013;26(6):1045-1057. doi:10.1007/s10278-013-9622-7

2.\ \ \ \ Bakas S, Akbari H, Sotiras A, et al. Advancing The Cancer Genome Atlas glioma MRI collections with expert segmentation labels and radiomic features. \textit{Sci data}. 2017;4:170117. doi:10.1038/sdata.2017.117

3.\ \ \ \ Menze BH, Jakab A, Bauer S, et al. The Multimodal Brain Tumor Image Segmentation Benchmark (BRATS). \textit{IEEE Trans Med Imaging}. 2015;34(10):1993-2024. doi:10.1109/TMI.2014.2377694

4.\ \ \ \ Baid U, Ghodasara S, Mohan S, et al. The RSNA-ASNR-MICCAI BraTS 2021 Benchmark on Brain Tumor Segmentation and Radiogenomic Classification. Published online July 5, 2021. http://arxiv.org/abs/2107.02314

5.\ \ \ \ Karschnia P, Young JS, Dono A, et al. Prognostic validation of a new classification system for extent of resection in glioblastoma: a report of the RANO resect group. \textit{Neuro Oncol}. Published online August 12, 2022. doi:10.1093/neuonc/noac193

6.\ \ \ \ Karschnia P, Vogelbaum MA, van den Bent M, et al. Evidence-based recommendations on categories for extent of resection in diffuse glioma. \textit{Eur J Cancer}. 2021;149:23-33. doi:10.1016/j.ejca.2021.03.002

7.\ \ \ \ Stupp R, Hegi ME, Mason WP, et al. Effects of radiotherapy with concomitant and adjuvant temozolomide versus radiotherapy alone on survival in glioblastoma in a randomised phase III study: 5-year analysis of the EORTC-NCIC trial. \textit{Lancet Oncol}. 2009;10(5):459-466. doi:10.1016/S1470-2045(09)70025-7

8.\ \ \ \ Ellingson BM, Wen PY, Cloughesy TF. Modified Criteria for Radiographic Response Assessment in Glioblastoma Clinical Trials. \textit{Neurotherapeutics}. 2017;14(2):307-320. doi:10.1007/s13311-016-0507-6

9.\ \ \ \ Li X, Morgan PS, Ashburner J, Smith J, Rorden C. The first step for neuroimaging data analysis: DICOM to NIfTI conversion. \textit{J Neurosci Methods}. 2016;264:47-56. doi:10.1016/j.jneumeth.2016.03.001

10.\ \ \ \ Rohlfing T, Zahr NM, Sullivan E V, Pfefferbaum A. The SRI24 multichannel atlas of normal adult human brain structure. \textit{Hum Brain Mapp}. 2010;31(5):798-819. doi:10.1002/hbm.20906

11.\ \ \ \ Jenkinson M, Smith S. A global optimisation method for robust affine registration of brain images. \textit{Med Image Anal}. 2001;5(2):143-156. doi:10.1016/s1361-8415(01)00036-6

12.\ \ \ \ Jenkinson M, Bannister P, Brady M, Smith S. Improved optimization for the robust and accurate linear registration and motion correction of brain images. \textit{Neuroimage}. 2002;17(2):825-841. doi:10.1016/s1053-8119(02)91132-8

13.\ \ \ \ Hoopes A, Mora JS, Dalca A V, Fischl B, Hoffmann M. SynthStrip: skull-stripping for any brain image. \textit{Neuroimage}. 2022;260:119474. doi:10.1016/j.neuroimage.2022.119474

14.\ \ \ \ Davatzikos C, Rathore S, Bakas S, et al. Cancer imaging phenomics toolkit: quantitative imaging analytics for precision diagnostics and predictive modeling of clinical outcome. \textit{J Med imaging (Bellingham, Wash)}. 2018;5(1):011018. doi:10.1117/1.JMI.5.1.011018

15.\ \ \ \ Kamnitsas K, Ledig C, Newcombe VFJ, et al. Efficient multi-scale 3D CNN with fully connected CRF for accurate brain lesion segmentation. \textit{Med Image Anal}. 2017;36:61-78. doi:10.1016/j.media.2016.10.004

16.\ \ \ \ Scarpace, L., Mikkelsen, T., Cha, S., Rao, S., Tekchandani, S., Gutman, D., Saltz, J. H., Erickson, B. J., Pedano, N., Flanders, A. E., Barnholtz-Sloan, J., Ostrom, Q., Barboriak, D., $\&$ Pierce LJ. The Cancer Genome Atlas Glioblastoma Multiforme Collection (TCGA-GBM) (Version 4) [Data set]. \textit{Cancer Imaging Arch}. Published online 2016. doi:https://doi.org/10.7937/K9/TCIA.2016.RNYFUYE9

17.\ \ \ \ National Cancer Institute Clinical Proteomic Tumor Analysis Consortium (CPTAC). The Clinical Proteomic Tumor Analysis Consortium Glioblastoma Multiforme Collection (CPTAC-GBM). \textit{Cancer Imaging Arch}. Published online 2018. doi:https://doi.org/10.7937/K9/TCIA.2018.3RJE41Q1

18.\ \ \ \ Mamonov AB KCJ. Data From QIN GBM Treatment Response. \textit{Cancer Imaging Arch}. Published online 2016. doi:10.7937/k9/tcia.2016.nQF4gpn2

19.\ \ \ \ Gerstner ER, Zhang Z, Fink JR, et al. ACRIN 6684: Assessment of Tumor Hypoxia in Newly Diagnosed Glioblastoma Using 18F-FMISO PET and MRI. \textit{Clin Cancer Res}. 2016;22(20):5079-5086. doi:10.1158/1078-0432.CCR-15-2529

20.\ \ \ \ Ratai EM, Zhang Z, Fink J, et al. ACRIN 6684: Multicenter, phase II assessment of tumor hypoxia in newly diagnosed glioblastoma using magnetic resonance spectroscopy. \textit{PLoS One}. 2018;13(6):e0198548. doi:10.1371/journal.pone.0198548

21.\ \ \ \ Puchalski RB, Shah N, Miller J, et al. An anatomic transcriptional atlas of human glioblastoma. \textit{Science (80- )}. 2018;360(6389):660-663. doi:10.1126/science.aaf2666

22.\ \ \ \ Bakas S, Sako C, Akbari H, et al. The University of Pennsylvania glioblastoma (UPenn-GBM) cohort: advanced MRI, clinical, genomics, $\&$ radiomics. \textit{Sci data}. 2022;9(1):453. doi:10.1038/s41597-022-01560-7

23.\ \ \ \ Calabrese E, Villanueva-Meyer JE, Rudie JD, et al. The University of California San Francisco Preoperative Diffuse Glioma MRI Dataset. \textit{Radiol Artif Intell}. 2022;4(6). doi:10.1148/ryai.220058

\begin{table}[H]
\begin{adjustbox}{max width=\textwidth}
\begin{tabular}{p{2.04cm}p{1.56cm}p{3.86cm}p{3.78cm}p{3.74cm}}
\hline
\multicolumn{5}{|p{14.98cm}|}{{\footnotesize \textbf{Supplementary Table 1. MRI acquisition parameters.}}} \\ 
\hline
\multicolumn{2}{|p{3.6cm}}{} & 
\multicolumn{1}{|p{1.56cm}}{\centering
{\footnotesize \textbf{Primary center}}} & 
\multicolumn{2}{|p{7.52cm}|}{\centering
{\footnotesize \textbf{Secondary center}}} \\ 
\hline
\multicolumn{2}{|p{3.6cm}}{\centering
{\footnotesize \textbf{Manufacturer, model, and Field strength}}} & 
\multicolumn{1}{|p{1.56cm}}{\centering
{\footnotesize General Electric, Signa HDxT, 1.5 T}} & 
\multicolumn{1}{|p{3.86cm}}{\centering
{\footnotesize Philips, Ingenia Ambition X, 1.5 T}} & 
\multicolumn{1}{|p{3.78cm}|}{\centering
{\footnotesize Philips, Achieva, 1.5 T}} \\ 
\hline
\multicolumn{2}{|p{3.6cm}}{\centering
{\footnotesize \textbf{Number of MRI studies}}} & 
\multicolumn{1}{|p{1.56cm}}{\centering
{\footnotesize 107 (89 $\%$)}} & 
\multicolumn{1}{|p{3.86cm}}{\centering
{\footnotesize 11 (9 $\%$)}} & 
\multicolumn{1}{|p{3.78cm}|}{\centering
{\footnotesize 2 (2 $\%$)}} \\ 
\hline
\multicolumn{1}{|p{2.04cm}}{\multirow{5}{*}{\parbox{2.04cm}{\centering
{\footnotesize \textbf{MRI sequence}}}}} & 
\multicolumn{1}{|p{1.56cm}}{\centering
{\footnotesize \textbf{T1ce}}} & 
\multicolumn{1}{|p{3.86cm}}{\centering
{\footnotesize TR/TE, 7.98 ms/2.57 ms; 3D; GRE; matrix, 512 x 512; slice thickness, 1 mm}} & 
\multicolumn{1}{|p{3.78cm}}{\centering
{\footnotesize TR/TE, 17.96 ms/6.43 ms; 3D ProSET; matrix, 230 x 230; slice thickness, 1 mm}} & 
\multicolumn{1}{|p{3.74cm}|}{\centering
{\footnotesize TR/TE, 25 ms/6.7 ms; ProSET, 3D; matrix, 256 x 256; slice thickness, 1.6 mm}} \\ 
\hhline{~----}
\multicolumn{1}{|p{2.04cm}}{} & 
\multicolumn{1}{|p{1.56cm}}{\centering
{\footnotesize \textbf{T1w}}} & 
\multicolumn{1}{|p{3.86cm}}{\centering
{\footnotesize TR/TE, 580 ms/7.56 ms; 2D;  FSE; matrix, 512 x 512; slice thickness, 5 mm}} & 
\multicolumn{1}{|p{3.78cm}}{\centering
{\footnotesize TR/TE, 525.6 ms/12 ms; 2D; SE; matrix, 228 x 227; slice thickness, 5 mm}} & 
\multicolumn{1}{|p{3.74cm}|}{\centering
{\footnotesize TR/TE, 456.2 ms/12 ms; 2D; SE; matrix, 249 x 191; slice thickness, 6 mm}} \\ 
\hhline{~----}
\multicolumn{1}{|p{2.04cm}}{} & 
\multicolumn{1}{|p{1.56cm}}{\centering
{\footnotesize \textbf{T2w}}} & 
\multicolumn{1}{|p{3.86cm}}{\centering
{\footnotesize TR/TE, 5220 ms/96.12 ms; 2D; FRSE; matrix, 512 x 512; slice thickness, 5 mm.}} & 
\multicolumn{1}{|p{3.78cm}}{\centering
{\footnotesize TR/TE, 5327.3 ms/110 ms; 2D; TSE; matrix, 232 x 232; slice thickness, 3 mm.}} & 
\multicolumn{1}{|p{3.74cm}|}{\centering
{\footnotesize TR/TE, 2456.2 ms/110 ms; 2D; TSE; matrix, 264 x 203; slice thickness, 5 mm.}} \\ 
\hhline{~----}
\multicolumn{1}{|p{2.04cm}}{} & 
\multicolumn{1}{|p{1.56cm}}{\centering
{\footnotesize \textbf{FLAIR}}} & 
\multicolumn{1}{|p{3.86cm}}{\centering
{\footnotesize TR/TE, 8002 ms/135.07 ms; 2D; FSE; matrix, 512 x 512; slice thickness, 4 mm}} & 
\multicolumn{1}{|p{3.78cm}}{\centering
{\footnotesize TR/TE, 5000 ms/375.8 ms; 3D; SPIR; matrix, 196 x 196; slice thickness, 1.2 mm}} & 
\multicolumn{1}{|p{3.74cm}|}{\centering
{\footnotesize TR/TE, 6000 ms/120 ms; 2D; FSE; matrix, 200 x 159; slice thickness, 2.8 mm}} \\ 
\hhline{~----}
\multicolumn{1}{|p{2.04cm}}{} & 
\multicolumn{1}{|p{1.56cm}}{\centering
{\footnotesize \textbf{DWI}}} & 
\multicolumn{1}{|p{3.86cm}}{\centering
{\footnotesize TR/TE, 8000 ms/111.7 ms; matrix, 128 x 160; slice thickness, 5 mm; b-values, 0 and 1000 s/mm2}} & 
\multicolumn{1}{|p{3.78cm}}{\centering
{\footnotesize TR/TE, 4600 ms/84.4 ms; matrix, 190 x 190; slice thickness, 5 mm; b-values, 0 and 1000 s/mm2}} & 
\multicolumn{1}{|p{3.74cm}|}{\centering
{\footnotesize TR/TE, 3414 ms/88.8 ms; matrix, 112 x 89; slice thickness, 5 mm; b-values, 0 and 1000 s/mm2}} \\ 
\hline
\multicolumn{5}{|p{14.98cm}|}{{\footnotesize T1ce $=$ contrast-enhanced T1w, T2w$=$ T2-weighted image, FLAIR $=$ Fluid-attenuated inversion recovery, DWI $=$ diffusion weighted image, TR $=$ repetition time, TE$=$ echo time, GRE $=$ gradient echo. FSE$=$ fast spin echo. FRFSE$=$ fast recovery fast spin echo. ProSET$=$ principle of selective excitation technique. SE$=$ spin echo. SPIR$=$ spectral presaturation with inversion recovery. TSE$=$ turbo spin echo. \par}} \\ 
\hline
\end{tabular}
\end{adjustbox}
\end{table}

\end{document}